\documentclass[prl,showpacs,preprintnumbers,amsmath,amssymb]{revtex4}

\usepackage[dvips]{graphics}

\begin{document}

\title{Are There Any Real Problems With Quantum Gravity?}

\author{Vlatko Vedral}
\affiliation{Clarendon Laboratory, University of Oxford, Parks Road, Oxford OX1 3PU, United Kingdom and\\Centre for Quantum Technologies, National University of Singapore, 3 Science Drive 2, Singapore 117543 and\\
Department of Physics, National University of Singapore, 2 Science Drive 3, Singapore 117542}

\date{\today}

\begin{abstract}
We present a short, general and accessible introduction to quantizing gravity in the Heisenberg picture. We then apply this formalism to the scenario where two spatially superposed masses interact through the gravitational field. We discuss some of the consequences for quantum gravity including the spin of the graviton, the notions of locality and causality, as well as going beyond the linear regime. We use the Schwinger action principle and the Heisenberg representation which we believe make the issues involved clearer. We conclude by commenting on a number of traditionally discussed apparent problems with quantizing gravity (only to, ultimately, deny their very existence).  
\end{abstract}


\maketitle                           %display desi d

\section{Introduction}

Heisenberg discovered an ingenious way to ``fix" classical physics. He decided to keep the classical laws of motion intact, which he chose to be governed by Hamilton's canonical equations. However, the entities that obeyed these equations were, according to Heisenberg, no longer just the ordinary numbers representing the relevant variables such as the position and momentum of a particle. They were instead what Dirac later called $q$ (for quantum)-numbers. A consequence of this was the famous Uncertainty Principle, namely that, because the $q$-numbers representing the position and momentum no longer commute (first fully appreciated by Born through the use of matrices), they could not be specified (or measured) to any desired accuracy. Planck's constant therefore acquired a new meaning as the limit of the simultaneous resolution of a particle's position and momentum.

The beauty of Heisenberg's logic is that it successfully applied to everything. You want to quantize the electromagnetic field? No problem. Keep Maxwell's equations, but upgrade the electric and the magnetic fields to $q$-numbers. It then automatically follows that certain pairs of components of the electric and the magnetic field cannot be simultaneously measured. There are, of course, immediate observable consequences of this, such as the existence of the quantum vacuum, the existence of photons, their indistinguishability, including the knock-on effects on matter such as the Lamb shift and the spontaneous emission of light from excited states. 

The ingenuity of Heisenberg's approach lies in the fact that, by quantizing things this way, you get to keep all the good stuff from classical physics. For instance, by keeping Maxwell's equations you retain all the desired properties of classical fields such as locality, causality and the general compliance with (special) relativity. It is just that the things subject to these relativistic symmetries are now $q$-numbers or operators. Heisenberg's idea was no doubt revolutionary, but the approach in physics has actually always been the same: keep the good stuff and modify/discard/upgrade the bad stuff.

This is not to say that there are no problems with Heisenberg's approach. For starters, which classical numbers do we decide to promote to $q$-numbers and how? We will come back to this question later, though the short answer is: it's work in progress. Ultimately, however, Heisenberg's approach (at least in ordinary quantum physics) is cumbersome from the computational point of view, which is how we justify the subsequent triumph of Schr\"odinger's way of doing it, namely the wave mechanics. 

But the spirit of wave mechanics is very different, even though the two approaches, the matrix and the wave mechanics, are fundamentally equivalent (though this equivalence has sometimes been contested). With Schr\"odinger we get the impression that the classical dynamical laws of motion have to be changed themselves. This is how we motivate the introduction of the Schr\"odinger equation and show that, in some special limit, it reduces to Newton's $F=ma$ (the Ehrenfest theorem).  

Schr\"odinger's approach to quantum physics also puts more emphasis on the wavefunction (misleading some people into believing that the wavefunction is the only reality). This famously led him to the concept of entanglement which, he said, was (not just one of many, but) the characteristic trait of quantum physics. A great strength of this is that it ultimately allows us to see quantum measurements as entanglements between different physical systems through the pioneering works of Mott, Everett and DeWitt among many others \cite{Zurek}. However, Schr\"odinger's approach, by demoting observers to mere ``other" quantum systems (which is a good feature!), downplays the role of observables. This, in turn, makes it less transparent regarding how and which of the good features of classical physics are maintained in quantum physics (not to mention the fact that observables and states have to be combined together into things we actually observe, irrespective of the approach we take to quantum physics). The common problems wave mechanics shares with Heisenberg is the decision regarding what to quantize and how exactly to do it. 

\section{Quantum Field Theory \`a la Heisenberg}

Here, we would like to discuss some aspects of quantum gravity, but from Heisenberg's perspective, since it seems that it is more suitable, especially when one deals with fields (which themselves are $q$-numbers). Two highly relevant descendants of Heisenberg in this case are Schwinger and K\"all\'en. Schwinger was the first person to be clearly dissatisfied by the fact that the whole approach to quantum physics is conceptually wrong. Namely, we start with classical physics first, then quantize. However, Schwinger complained, given that quantum physics is an exponentially larger arena, shouldn't we first start with quantum physics and then derive classical physics in some special limit? 

Schwinger's answer was: the quantum action principle \cite{Schwinger}. It postulates a constraint on the variation of a fundamentally quantum entity, the transition amplitude, in terms of changes of the corresponding relevant $q$-numbers. There is no classical physics in sight (other than the background spacetime!). Plus, it suffices to give us the correct equations of motion as well as the corresponding commutation relations. Let us illustrate the concept with a single particle where the quantum action principle has the following form:
\begin{equation}
\delta \langle b,t_2|a,t_1\rangle = \langle b,t_2| \delta L  |a,t_1\rangle = \langle b,t_2|(p\delta x - H\delta t)|^2_1 |a,t_1\rangle \;\; ,
\end{equation}
where $L=\int_1^2 p\delta x - H\delta t$ is the Lagrangian. In other words, the variation of the amplitude between two points $1$ and $2$ only depends on the variation at the boundaries. This can only be because the variation along the path connecting $1$ and $2$ is assumed to vanish (which is what gives us the Heisenberg equations of motion for the operators $x$ and $p$). But the interesting point about Schwinger's action principle is that it not only contains the dynamics, but it also leads to the correct commutation relations between $x$ and $p$. In that sense it almost looks as though we get more out of it than what we put into it. This is because the boundary terms contain the relevant quantum generators.

To see the latter, we note that the variation, as always in quantum physics, is assumed to be unitary, such that $U=1-i/\hbar \delta L$. An observable $O$, then evolves according to: $O\rightarrow U^{\dagger}OU = O -i/\hbar [O,\delta L]$. Let us assume only the variation in position (and not in time), i.e. $\delta L = p\delta x$. Then, substituting $O=x$, we obtain $\delta x = -i/\hbar [x,p\delta x]$. This is possible if $[x,p] = i\hbar$ (though this approach does not rule out more general commutation relations). This can be applied to any generators including temporal shifts, rotations and boosts. 

The same can be done for fields, and it is instructive to now use Schwinger's method within the context of the electromagnetic field itself. We will ultimately model gravity in exactly the same way. The free EM field Lagrangian is given by $L=-1/(4c)\int F_{\mu\nu}F^{\mu\nu} \sqrt{-g}d\Omega$, where $F_{\mu\nu}$ is the electromagnetic field Faraday tensor, $g$ is the determinant of the metric tensor and $\Omega$ is the $4$ dimensional spacetime volume of integration. The variation of $L$ contains the bulk term which leads to Maxwell's equations, as well as the surface (boundary) term of the form $-1/c \oint F^{\mu\nu} A_\nu \sqrt{-g} df_\mu$, where $A_{\nu}$ is the vector potential and $df_\mu=\delta_{0\mu} dV$ is the element of the hypersurface over $\Omega$. The same logic as before leads us to conclude that the only non-trivial commutator (when $F$ is treated as an operator and therefore so are the $A$, $E$ and $B$ fields) is $[A_\alpha (x,t),E^\beta (x',t)]=i\hbar c \delta_{\beta\alpha} \delta (x-x')/\sqrt{\gamma}$ where $\gamma$ is the determinant of the spatial part of $g_{\alpha\beta}$ and $E^\beta = F^{0\beta}$ is the electric field part of the tensor. From here on, we can also derive the commutation relations for the electric and the magnetic field components. 

The quantum action principle therefore leads to the correct dynamics, i.e. the correct Heisenberg equations of motion, as well as the correct commutation relations - in terms of the compatibility, as we mentioned before - whenever we have the appropriate form of the Lagrangian. With Schwinger, Heisenberg's approach to quantum field theory is set on a firmer logical foundation. It is also more compact than other approaches. 

The climax is reached with the work of K\"all\'en. K\"all\'en's idea was to phrase quantum electrodynamics in the fully non-perturbative fashion in the Heisenberg picture. 
Here, of course, we have the equations of motion for the fermionic field, as well as for the electromagnetic field (both of which can be derived from the QED Lagrangian), which couple to one another (as in Maxwell's equations). Without repeating the whole variational procedure, the equations of motion can be written as \cite{Kallen}:
\begin{eqnarray}
\psi (x) & = & \psi^{(0)} (x) - ie\int S_R (x-x') \gamma A(x')\psi (x') dx' \\
A_\mu(x) & = & A^{(0)}_\mu (x) +\frac{ie}{2}\int D_R (x-x') [\bar\psi (x'),\gamma_\mu \psi (x')] dx' \; ,
\end{eqnarray} 
where $\psi (x)$ is the usual Dirac field and $S_R (x-x')$ and $D_R (x-x')$ are the relevant retarded propagators for the fermionic and the EM field respectively (the bar denotes the conjugate). The exact form of propagators is known in terms of Bessel functions, but is not of interest to us here (one of them will be explicitly presented below). The operators $\psi^{(0)} (x)$ and $A^{(0)}_\mu (x)$ are the solutions of the free-field equations:
$\big(\gamma \frac{\partial}{\partial x} +m \big)  \psi^{(0)} (x)  =  0$ and
$\Box A^{(0)}_\mu (x)  =   0$. They constitute what one might call the initial conditions. After that, the interaction is ``turned on". 

The above field equations are the same as in classical electrodynamics, but they involve quantum operators representing the respective fields. It is clear what the classical electromagnetic field is, but by the term ``classical fermionic field equations" we here mean the Dirac equation. This is logically at the same level of description as the classical Maxwell's equations; otherwise, there are no other classical fermionic field equations \cite{Schweber,Vedral-class-AB}. 

In almost all practical applications the coupled equations have to be solved perturbatively. This is unfortunately an unavoidable feature of all field theories, classical or quantum. Despite having to solve the equations for the Dirac field (which involve the fermionic field operator coupling to the vector potential) we can always express the local observables in terms of the local current operator $j_\mu (x) = -e [\bar\psi (x),\gamma_\mu \psi (x)]$ and the local vector potential operator $A_\mu (x)$ (here $\bar\psi (x)$ has its usual meaning). These are both Hermitian operators and therefore represent (locally defined) observables (though one has to be careful to take into account the gauge invariance). 

To compute the next state of the current and the vector potential we use the field equations perturbatively (by ``plugging the equations back into themselves" recursively). Namely, the $q$-numbers at step $1$ of iteration are a function of the $q$-numbers at step $0$, as follows:
\begin{eqnarray}
j^{(1)}_\mu (x) & = & j^{(0)}_\mu + \frac{1}{2} \int [\bar\psi^{(0)} (x), \gamma_\mu S_R (x-x')\gamma_\nu \psi^{(0)} (x')]  dx' + \int [\bar\psi^{(0)} (x) \gamma_\nu S_A (x'-x),\gamma_\mu \psi^{(0)} (x')] A^{(0)}_\nu(x') dx'\\
A^{(1)}_\mu(x) & = & A^{(0)}_\mu(x) + \frac{ie}{2}\int D_R (x-x') [\bar\psi^{(0)} (x'),\gamma_\mu \psi^{(0)} (x')] dx'
\end{eqnarray} 
and, by continuing the procedure, we can obtain an arbitrarily high order of accuracy. In every subsequent iteration we update the relevant operators through interactions sampled from points that can causally and locally affect the operator at the point of interest (the retarded propagators vanish for space-like separated points, which is know as the condition of ``micro-causality"). We therefore preserve all the Lorentz symmetries of the corresponding classical fields, while at the same time we still deal with the quantum operators that represent the relevant physical observables (which are ultimately experimentally accessible, see \cite{Scharf}).  The author has recently applied the Heisenberg picture to the quantum double slit experiment to show that there is no non-locality present in this description, which is fairly obvious from the above treatment \cite{Vedral-double-slit}. A more complicated scenario involving electronic interference in the electromagnetic field can be found in \cite{Nicetu}. 

\section{Quantum Gravity}

Two of Schwinger's students, Arnowitt and Deser \cite{Arnowitt}, and independently Schwinger himself (in an unpublished manuscript), applied this kind of logic to quantum gravity in the linear regime. There is then an argument to be made that the higher order terms could be obtained by successive iterations much as in the case of QED. In fact, by reversing the logic in a way that would certainly please Schwinger, we can start with gravity as a quantum field and derive the classical GR from it. This follows an idea of Thirring's \cite{Thirring} and was developed by Boulware and Deser (Feynman's work belongs here too \cite{Feynman}).

For the sake of completeness, we first summarise briefly the Schwinger action principle as applied to General Relativity. The surface term when minimizing the variation of the Einstein-Hilbert Lagrangian can be shown to be $\oint X^{\mu} df_\mu$, where $X^{\mu} = g^{\rho\nu}\delta \Gamma^\mu_{\rho\nu} - g^{\mu\nu}\delta \Gamma^\rho_{\nu\rho}$. Written symbolically the equation for $\Gamma$ is $\delta \Gamma \propto [\Gamma, g\Gamma]$ from which it follows that the commutator $[\Gamma^0_{\rho\nu},g_{\alpha\beta}]$ cannot be zero in general. This tells us that the affine connection elements $\Gamma$ are like the electromagnetic field components, while the metric $g$ is like the vector potential. Needless to say, we could also talk about different components of $\Gamma$ as non-commuting. We will come back to a concrete example of the non-commuting gravitational components in the next section (for a more esoteric example in the Einstein-Cartan gravity, see e.g. \cite{Poplawski}). Many people realised that this kind of non-commutativity in gravity must follow logically, if one applies the usual quantum field theory, starting from probably Bronstein (for a detailed account see \cite{Bronstein}), then including Peres and Rosen \cite{Peres} as well as Regge \cite{Regge}. 

This would suggest that the quantization of gravity ought to be as straightforward as the quantization of light. However, there is one important difference between the electro-dynamics and the geometro-dynamics which was first clearly appreciated by Bronstein, and this was well before anyone else entered the game. In GR there seems to be a fundamental limit on size imposed by black holes, that has to somehow be combined with the limitations imposed by the Heisenberg Uncertainty Principle as understood to apply to the different components of the affine connection. In that sense, as Bronstein expected, even the individual components of the gravitational field may not be measurable infinitely accurately. The reason is that, in order to measure the field strength, we need a test mass (something that responds to the field depending on its strength), and a higher and higher accuracy requires us to increase the mass (or energy) of the probe more and more. Increasing the precision means confining this mass to a smaller and smaller volume, which ultimately necessarily will exceed the Schwarzschield limit. We don't quite know what happens here (classically nothing can escape the black hole, however, it is not clear at all how to describe what would happen to a quantum black hole). 

Despite this difference between electrodynamics and GR, it is possible to have a perturbative approach to quantum gravity that is finite at each order of perturbation \cite{Scharf}. This can be done by using the method of Epstein and Glaser which has the advantage that it uses mathematically well-defined entities, namely free asymptotic fields. These can be defined in a ``smeared" way such that they are Lorentz covariant as well as gauge invariant in which case all the ultraviolet divergencies are avoided. 

The simple reason why there is no such limit on the individual electric and magnetic field components is that the electrical charge couples to the field in a way independent of the mass \cite{Bohr,Heitler}. As far as gravity, mass is always also a gravitational charge since everything with energy is seen by gravity. As we said, it remains to be seen what becomes of black holes in the full theory of quantum gravity, but as far as the low energy limits of things is concerned, it seems that there are no fundamental issues to be resolved. This is the limit to which we fully turn now. 

\section{Entanglement through gravity and all that}

We would now like to apply a K\"all\'en-like method to a problem in linear quantum gravity of two interacting massive particles each in a spatial superposition of two modes \cite{MAVE,MAVE2,SOUG}. The reason we do this is to demonstrate that this is nothing but a slightly more involved case of the quantum double slit experiment (as Feynman famously said, the double slit experiment presents the only mystery in quantum physics). Related proposals \cite{Howl,Rijavec} can be tackled the same way.

In the Newtonian limit, the Einstein field equations assume the form:
\begin{equation}
\Box h_{\mu\nu} = -16\pi G (T_{\mu\nu}-\frac{1}{2}\eta_{\mu\nu}T)
\end{equation}
where $h_{\mu\nu}$ is the perturbation of $g_{\mu\nu}$ away from the flat Minkowski metric $\delta_{\mu\nu}$. The solution of the above equation is: 
\begin{equation}
h_{\mu\nu} (x) = -16\pi G \int D_R (x-x') (T_{\mu\nu} (x')-\frac{1}{2}\eta_{\mu\nu} (x') T (x'))d\Omega \;\; ,
\end{equation}
which is just a simple instance of the retarded solution as in the case of the electromagnetic field \cite{Rohrlich}. 

It is now simple to proceed. If the masses are moving slowly, $T_{00}=\rho$ is the only non-vanishing component of $T_{\mu\nu}$. Then we have that
\begin{equation}
h_{00} (x) = -8\pi G \int \frac{\rho (t-|x-x'|))}{|x-x'|} dV
\end{equation}
which has twice the value of Newton's potential (the factor of $2$ is the appropriate normalisation). However, unlike Newton's gravity, this formula is relativistically causal, as ought to be the case (that was the whole point of introducing fields into physics). We can now easily quantize this, which simply means upgrading both 
$h_{00}$ and $\rho$ into $q$-numbers. The quantum equations becomes
\begin{equation}
\hat h_{00} (x) = -8\pi G \int \frac{\hat b^{\dagger}\hat b (t-|x-x'|))}{|x-x'|} \rho dV
\end{equation}
where $\hat b^{\dagger}\hat b$ is the number operator for the mass generating the field at $x$, which now itself is quantized.  This is the field theoretic quantized version of the Newtonian gravitational potential. 

In order to compute the phase in the double mass interference experiment, 
we can first calculate the $h_{00}$ generated by one mass and then couple that to the other mass via $H=-h^1_{00}T^2_{00}$ where the superscripts correspond to the two masses. The situation is completely symmetric with respect to the two masses and the resulting state ends up being 
\begin{equation}
\exp\left[ i\int \hat h^1_{00} (x) \hat T^2_{00} (x)dV t\right] |\Psi_0\rangle \;\;  ,
\end{equation}
where $|\Psi_0\rangle$ is the initial state of the superposed masses. This clearly leads to the maximally entangled state as predicted by Bose and Marletto and the author \cite{MAVE,SOUG}. 

Two observations are in order. GR gives us the Newtonian phases in the lowest order, but also presents a causal explanation regarding how they are established. We therefore avoid the Newtonian instantaneous action at a distance, this time with gravitational $q$-numbers. Secondly, it is clear that only the $h_{00}$ component is relevant and that therefore the scalar gravity (scalar meaning just one component) suffices to explain the entanglement between the masses. In other words, spin-$0$ quanta of the gravitational field will do the job (possibly spin $1$ might be needed if we accept that the masses can never be entirely stationary and therefore their momentum components become relevant too; this is something that lies within the domain of gravito-magnetism, but would take us beyond the intent of this paper - see \cite{Mashhoon}). 

We mention in passing that the same is true of the static Coulomb interaction in quantum electrodynamics. It too is mediated by the scalar mode of the vector potential (i.e. its time-like component) in the Lorenz gauge. The gauge issues are even more complicated in GR, and they present us with a challenge when we quantize because of the drastic effect that different gauges have on how we partition the field into subsystems. We will not comment on this here and the interested reader is referred to the excellent exposition in \cite{Scharf}. 

One can naturally ask what would convince us that gravity is actually mediated by gravitons which are quanta of spin $2$ (this follows, speaking somewhat loosely, from the fact that $g_{\mu\nu}$ is a two index tensor and a spin $1$ is required for quantization of each of the indices). The answer is that the simplest thing to do, but by no means the only one, is to involve the electromagentic field. The reason is that the EM field is represented by the Faraday tensor which is in GR part of $T_{\mu\nu}$. Therefore, when the electromagnetic field generates $h_{\mu\nu}$, its other components will now also become relevant (I know that this is a purely mathematical argument, but this is all we have at present). The situation is the same as what happens in the bending of light in the conventional classical GR where the $h_{00}$ component is, on its own, insufficient to explain the observed result (underestimating it by the famous factor of $2$). An experiment would therefore be along the lines of interacting a light beam with a superposed massive object. In that case the light beam would bend differently depending on the centre of mass of the superposed massive object and the path of light would become entangled with the centre of mass of the massive object. 

It seems to us that even though this experiment is in principle possible, it is much more beyond the experimental reach than the original Bose-Marletto-Vedral proposal. The magnitude of the effect is of the second order (the Newtonian limit cannot fully account for these genuinely GR effects) and the equation for the metric we need to use is:
\begin{equation}
\Box h_{\mu\nu} = -16\pi G (\partial_\lambda h_{\mu\alpha} \partial^\lambda h_{\alpha\nu}) \;\; ,
\end{equation}
reflecting the fact that the energy of the gravitational field is now part of the energy momentum tensor ($T_{\mu\nu}\propto \partial_\lambda h_{\mu\alpha} \partial^\lambda h_{\alpha\nu}$), i.e. gravity indeed gravitates (see \cite{Biswas} for the derivation of the Mercury perihelion and light bending from the above equation). This equation can also be solved iteratively just as was done in the case of QED \cite{Scharf}. We will make comments about the experimental feasibility of different orders in the conclusions. 

A final comment is that various semi-classical, so-called hybrid, models have been proposed as a way of preserving the classicality of gravity while allowing for the masses to still become entangled. This is simply impossible. Together with Marletto, I have made a more general argument regarding quantization of gravity that goes beyond having to assume the above equations of motion \cite{MAVE,MV-njp,MV-njpq,MV-PRD}. It only relies on the locality of the interaction and the fact that the masses only couple through gravity. In this case, however, because the argument is general and applicable to any kind of a mediator, it is hard to tell which gravitational properties end up being non-commuting.  All classical models of the gravitational field are therefore refuted by the creation of entanglement \cite{MV-when,Marconato}. But what are the non-commuting degrees of freedom that force gravity to be quantum if we assume GR in the lowest Newtonian approximation? 

We recall the commutator derived from Schwinger's action principle is of the form $[\Gamma^0_{\rho\nu},g_{\alpha\beta}]\propto \delta_{\rho\alpha}\delta_{\nu\beta} + \delta_{\rho\beta}\delta_{\nu\alpha} - \delta_{\rho\nu}\delta_{\alpha\beta}$. Now, in the Newtonian regime, the only commutator surviving is $[\Gamma^0_{00},g_{00}]\neq 0$. This implies that $[\partial_0 h_{00}, h_{00}]\neq 0$.  This simply means that the operator representing the Newtonian potential $\phi (x)$, and its conjugate (represented by its temporal derivative) do not commute, i.e. $[\partial_t \phi (x), \phi (x')]\propto \delta (x-x')$. They can be represented by the same combination of creation and annihilation operators as the position and momentum $q$-numbers and both appear in the gravitational Hamiltonian which then couples to the matter. Therefore, the BMV experiment, if the linear quantum gravity model is correct as outlined above, already proves the complementarity between one component of the affine connection and one component of the metric tensor. To test the non-commutativity of other components we would have to perform more intricate experiments as the ones involving the light field or some higher order effects in the BMV experiment with two material particles. Even though our exposition follwed a particular way of thinking about qunatum gravity, we believe that all other fully quantum approaches would be in agreement about the low energy limit. 

As a side remark, it is interesting to note one general point pertaining to any superposition experiment in the gravitational field. In the linear gravitational regime, the phases we obtain have the form $\xi = mc^2\tau/\hbar$. The relative phase between different gravitational fields could therefore be though of in $3$ different ways. One way is in the spirit of the dictum ``gravity is the flow of time from space to space", which says that the relative phase $\delta \xi = mc^2\delta \tau/\hbar$, where $\delta \tau/ = gh/c^2$ is the redshift between two points in Earth's gravity at heights separated by $h$. The second way is to think that the mass varies from point to point, so that $\delta \xi = \delta mc^2\tau/\hbar$ where $\delta m = mgh/c^2$. In other words, the bare mass gets dressed by the local gravitational field, which accounts for the relative phase between two locations (this is reminiscent of the mass renormalisation in the EM field). The third way is to think of the speed of light being different at two different heights, i.e. $\delta \xi = 2mc(\delta c) \tau/\hbar$, where now $\delta c = gh/2c$ \cite{Dicke}. 

The three accounts all lead to the same relative phase and are therefore physically indistinguishable at this level. The only conclusion that we can safely draw is that the product $mc^2\tau$, which has the units of action, gets modified due to gravity. Perahps this adds extra weight to Schwinger's strategy that quantum action is ultimately the fundamental entity in terms of which everything else ought to be phrased? Only time will tell. 

\section{Conclusions}

We believe that by approaching the problem of quantum gravity the Heisenberg way, we compromise the least on clarity and transparency. But, we also can account for all near-term experimentally accessible observations. It seems that this approach has no conceptual difficulties over and above the usual issues in any other quantum theory of fields \cite{Weinberg,Kallen,Schweber,Scharf}. We, in fact, doubt the genuine existence of a number of problems that are frequently flagged up in the literature when it comes to quantizing gravity. Summarizing, our view is that there is:
\begin{itemize}
\item {\bf no problem with the Bell type non-locality and gravity}. Clearly the formulation of quantized gravity can be made to be causal and could be designed to handle entanglement between matter or radiation and the gravitational field as well as having entanglement within the gravitational field itself. Entanglement is, in fact, a necessary feature of quantum gravity as it is with all other quantum systems. The key to any of this is to think of $q$-numbers as the underlying elements of reality which in this case are the components of $g$ and $\Gamma$ \cite{Vedral-local} and to understand entanglement through correlations between these $q$-numbers;
\item {\bf no problem with the equivalence principle}. This principle, like other classical principles resulting from various symmetries, can be phrased perfectly meaningfully with $q$-numbers. When quantum superpositions comply with classical principles, such as the equivalence principle, or the energy conservation principle, this is so in a branch by branch way (meaning that each element of the superposition behaves like the corresponding classical world) \cite{MV-tot,MV-sagn}.  In fact, Weinberg \cite{Weinberg} turned the whole thing round and argued, within the scattering matrix formalism, that the equivalence principle is a consequence of quantum physics (more precisely of unitarity and Lorentz invariance in the case of a mediator with spin $2$). It is exactly the energy-momentum conservation that forces all fields to couple to gravity equally. 
\item {\bf no non-local observables, or gauge/diffeomorphism problems}. All observables are local (as in our treatment of the gravitationally induced entanglement between two masses and as in all other equations above), and can be calculated in any chosen frame. One can, as in the case of the EM field, use holonomies as observables (which is one way of making the gauge invariance manifest), but that does not mean that the fields are intrinsically non-local (see my recent work with Marletto \cite{MV-AB}). Any description can be made non-local (by e.g. removing the field) while still complying with microcausality (as in the Wheeler-Feynman absorber theory \cite{Hoyle}), so this is clearly just a generic mathematical property;
\item {\bf no problems with non-linearities, of the ``gravity-gravitates" type} (i.e. because of the fact that gravitational field carries energy and is, therefore, itself subject to gravity). All field theories with interactions are actually non-linear \cite{Weinberg-PRL} which presents us with huge calculational difficulties, but none of them are fundamentally problematic due to this reason. Of course, there may well be a better way to handle non-linearities and possibly do non-perturbative treatments, but at present we are stuck with the method that works best;
\item {\bf no measurement problem}. This, in my view, is self-evident if everything is treated quantum mechanically. Any measurement is just the establishment of correlations between $q$-numbers of one system and $q$-numbers of another system through an interaction that is described by a unitary transformation \cite{Vedral-local}. Von Neumann already treated measurements this way in order to avoid the need for an abrupt collapse. Observers themselves are, according to this, just other quantum systems and need not possess any special additional properties. In the above case, when masses couple to the field, either can be said to measure the other. The perfect symmetry between entangled subsystems is the ultimate reason behind abolishing the dichotomy between the observers and the observed.  
\end{itemize}
I felt the need to spell out the above denials of problems (the list is actually much longer) simply because it seems to me that much of the literature is concerned with one or more of them in completely unproductive ways. Thinking of the above issues as problematic (more frequently than not) leads us into blind alleys out of which we will never come (to paraphrase Feynman, who used a similar expression in the context of the double-slit experiment). Rather, we need to take both quantum physics and GR seriously and think creatively how to test them in the domains where both are relevant. 

While all this might suggests to the reader that quantum gravity is no worse than any quantum field theory, it does not say anything about whether quantum field theory itself may be troublesome (when applied to any field). The occurrence of infinities, for instance, may be symptomatic of this failure of quantized fields, irrespective of whether they could be renormalised or not \cite{Scharf}. It is also perfectly possible that the quantum uncertainty in the metric tensor prevents us from dividing points into space-like, time like and null separated at some level of resolution. If two events could be in a superposition of space-like, time-like or null, then perhaps it is wrong to impose the condition known as microcausality the way it is conventionally done. Perhaps it is also possible that spacelike separated fields, represented as $q$-numbers, ought not to commute (as in the non-commutative geometry). In that sense, how should one quantize the canonical variables in the view of the indefinite causal orders in spacetime? 

At present, we sadly lack the relevant experimental evidence to decide on how to proceed in this domain. The perturbative quantum gravity \cite{Woodard}, whose basic tenets may well be tested in the near future in the lowest order, looks set to be confirmed. Then we will at least know that some aspects of gravity are quantum. The question is, what happens next? Will the lowest order GR contribution also be confirmed as affecting the quantum superposition? Yes, it is easy now to measure the gravitational time dilation with our atomic clocks (half a meter shift in height in Earth's gravitational field on the surface corresponds to the $1$ part in $10^{16}$ change in clock's frequency), but it seems hard to make a superposition of an atomic clock across two heights. What about the lowest order effects of the quantum gravitational vacuum? This is much smaller, and seems well beyond the experimental reach. Ditto for detecting gravitons (the same way we detect photons). 

So, the good news is that all seems fine with quantum gravity in the standard Heisenberg field theoretic approach. We can use the lowest order to experimentally confirm that gravity must be quantum. Higher order perturbations can be consistently calculated too, at least in principle \cite{Scharf,Bern}. The bad news might be that we may never be able to test most of them. Unless, of course, some observations from the early universe, or extreme astronomical objects like black holes and neutron stars come to rescue. 

\textit{Acknowledgments}: The author thanks Chiara Marletto on extensive discussions of related topics. This research has been supported by the National Research Foundation and the Ministry of Education, Singapore, as administered by Centre for Quantum Technologies, National University of Singapore.

\end{document}